\documentclass[12pt]{article}
\usepackage{amssymb}
\usepackage{amsmath}
\newtheorem{proposition}{Proposition}
\begin{document}
\title{Supersymmetric Two-Boson Equation: \\Bilinearization and Solutions}
\author{Q.\ P.\  Liu  and Xiao-Xia Yang \\ \em\small Department of Mathematics\\
\small \em China University of Mining and Technology\\
\small\em Beijing 100083, China }

\date{} \maketitle
\begin{abstract}
A bilinear formulation for the supersymmetric two-boson equation is
derived. As applications, some solutions are calculated for it. We
also construct a bilinear B\"{a}cklund transformation.
\end{abstract}

\section{Introduction}
The whole theory of solitons originated from a single partial
differential equation -- the celebrated Korteweg-de Vries (KdV)
equation, which is regarded as one of the most important systems
in mathematics and physics. A supersymmetric extension of the KdV
equation was introduced by Manin and Radul \cite{mr} (see also
\cite{mathieu}). Much research has been conducted on this $N=1$
supersymmetric KdV system and many remarkable properties have been
established. Here we mention just a few, bi-Hamiltonian
structures\cite{op}, Painlev\'{e} property\cite{ma}, infinite many
symmetries, Darboux transformation \cite{liu} and B\"{a}cklund
transformation (BT) \cite{lx} and bilinear forms
\cite{my}\cite{ca}\cite{cgr}.

Another equally important or more remarkable system is the
Broer-Kaup (BK) system \cite{broer}\cite{kaup}, which was solved
by the inverse scattering transformation and certain soliton
solutions are found. We remark here that both the classical
Boussinesq (cB) system and the dispersive water wave (DWW)
equation are equivalent to the BK equation. Kupershmidt in
\cite{Kup} constructed three local Hamiltonian structures for the
BK or DWW system and its B\"{a}cklund and Darboux transformations
are studied in \cite{li}. This system has various interesting
solutions, such as standard solitons \cite{kaup}\cite{hs}, fusion
and fission solitions \cite{mm}\cite{skm} and rational solutions
\cite{nh}. We also mention that the two-boson equation is one more
name for this system. It is shown that this equation appears in
matrix model theory \cite{aratyn}\cite{tb}.

A supersymmetric version of this system is proposed by Brunelli and
Das \cite{bd} and reads as
\begin{equation}\label{stb}
\left\{\begin{tabular}{l} $\Phi_{0,t}=-({\cal D}^4\Phi_0)+({\cal
D}({\cal
D}\Phi_0)^2)+2({\cal D}^2\Phi_1)$, \\[12pt]
$\Phi_{1,t}=({\cal D}^4\Phi_1)+2({\cal D}^2(({\cal
D}\Phi_0)\Phi_1)),$
\end{tabular}\right.
\end{equation}
where $\Phi_0$ and $\Phi_1$ are fermionic superfields depending on
usual independent variables $x$ and $t$ and the Grassmann variable
$\theta$. ${\cal D}=\partial_{\theta}+\theta\partial$ is the usual
super derivative. The system is shown to be a bi-Hamiltonian
system, have a Lax representation and various reductions
\cite{bd}.

The purpose of the present Note is to study the supersymmetric two
boson (sTB) from the viewpoint of solutions. We will show that the
system can be casted into Hirota's bilinear form and this in turn
provides us a way to find solutions.

The Note is organized as follows. In the next section, we will
transform the sTB equation into bilinear form. Then section 3 will
be devoted the construction of solutions. In section 4, we
construct a bilinear B\"{a}ckulund transformation for the sTB
system. Final section contains our discussion and conclusion.

\section{Bilinear Form}
To obtain a bilinearization of the sTB system (\ref{stb}),  we first
reformulate it. Let
\[
u={\cal D} \Phi_0,\;\; \alpha=\Phi_1
\]
then the system  (\ref{stb}) is transformed into
\begin{equation}\label{stb1}
\left\{\begin{tabular}{l}
$u_t=-u_{xx}+2uu_x+2({\cal D}\alpha_x)$, \\[12pt]
$\alpha_t=\alpha_{xx}+2(u\alpha)_x$.
\end{tabular}\right.
\end{equation}

Now we introduce the following transformations for the dependent
variables
\begin{equation}\label{sub}
u=\left(\ln{\tau_2\over \tau_1}\right)_x,\quad \alpha=({\cal
D}\ln{\tau_2})_x,
\end{equation}
substituting above expressions into the first equation of
(\ref{stb1}), we obtain
\[
\left[{1\over
\tau_1\tau_2}(\tau_{1,t}\tau_2-\tau_1\tau_{2,t}+\tau_{1,xx}\tau_2-2\tau_{1,x}\tau_{2,x}+\tau_1\tau_{2,xx})
\right]_x=0
\]
which gives us
\begin{equation}\label{bi1}
 (D_t+D_x^2)\tau_1\cdot \tau_2=0
\end{equation}
where the Hirota derivative is defined as
\[
D_t^mD_x^n f\cdot g=\left({\partial\over\partial
t_1}-{\partial\over\partial
t_2}\right)^m\left({\partial\over\partial
x_1}-{\partial\over\partial x_2}\right)^n
f(x_1,t_1,\theta_1)g(x_2,t_2,\theta_2)\left|_{\substack{x_1=x_2\\
t_1=t_2}}\right..
\]

Let us now turn to the second equation of (\ref{stb1}). After
substituting (\ref{sub}) into it and some calculations, we obtain
\begin{eqnarray}\label{5}
\left\{\lefteqn{{({\cal D}\tau_2)\over
\tau_1\tau_2^2}\left[\tau_1\tau_{2,t}-\tau_1\tau_{2,xx}+2\tau_{1,x}\tau_{2,x}\right]{}}\right.\nonumber\\
&& \left. {}-{1\over \tau_1\tau_2}\left[\tau_1({\cal
D}\tau_{2,t})-\tau_1({\cal D}\tau_{2,xx})+2\tau_{1,x}({\cal
D}\tau_{2,x})\right]\right\}_x=0.
\end{eqnarray}
It is interesting that the left hand side of the equation
(\ref{5}) can be represented as
\[
\left\{{1\over 2\tau_1\tau_2}\left[{\cal D}((D_t+D_x^2)\tau_1\cdot
\tau_2)-(SD_t+SD_x^2)(\tau_1\cdot \tau_2)\right]-{({\cal
D}\tau_2)\over \tau_1\tau_2^2}\left[(D_t+D_x^2)\tau_1\cdot
\tau_2\right]\right\}_x
\]
therefore, taking account of the equation (\ref{bi1}) we obatin
the other bilinear equation
\begin{equation}\label{bi2}
\quad (SD_t+SD_x^2)\tau_1\cdot \tau_2=0
\end{equation}
where we used a super version of Hirota derivative which is
introduced in \cite{my} \cite{ca}. Its definition is following
\begin{eqnarray*}
\lefteqn{SD_t^m D_x^n f\cdot g=} \\
&&({\cal D}_{\theta_1}-{\cal
D}_{\theta_2})\left({\partial\over\partial
t_1}-{\partial\over\partial
t_2}\right)^m\left({\partial\over\partial
x_1}-{\partial\over\partial x_2}\right)^n
f(x_1,t_1,\theta_1)g(x_2,t_2,\theta_2)\left|_{\substack{x_1=x_2\\
t_1=t_2\\ \theta_1=\theta_2}}\right..
\end{eqnarray*}

Thus we have succeeded in converting the sTB system (\ref{stb})
into Hirota's bilinear form. For convenience, we collect them
below
\begin{equation}\label{BI}
\left\{ \begin{tabular}{l}
$(D_t+D_x^2)\tau_1\cdot \tau_2=0$,\\[12pt]
$(SD_t+SD_x^2)\tau_1\cdot \tau_2=0$.
\end{tabular}\right.
\end{equation}

\section{Solutions}
It is well known that Hirota's bilinear form is ideal for
constructing interesting particular solutions. Next we shall show
that a class of solutions can be calculated for the sTB equation.
To this end, let
\[
\tau_1=\epsilon f,\;\; \tau_2=1+\epsilon g_1+\epsilon^2
g_2+\epsilon^3g_3+\cdots,
\]
substituting above expressions into the bilinear equations
(\ref{BI}) and collecting the alike power terms, we have
\begin{equation}\label{b1}
\epsilon^1: \;\; (D_t+D_x^2)(f\cdot 1)=0, \;\;
(SD_t+SD_x^2)(f\cdot 1)=0,
\end{equation}
and for $i>1$
\begin{equation}
\epsilon^i: \;\;(D_t+D_x^2)(f\cdot g_i)=0, \;\;
(SD_t+SD_x^2)(f\cdot g_i)=0 \label{b2}.
\end{equation}
From the equations (\ref{b1}), we have
\[
f_t+f_{xx}=0,\;\; {\cal D}(f_t+f_{xx})=0.
\]
Thus we may take
\[
f=\mbox{e}^{kx-k^2t+\theta\xi}
\]
where $k$ is an ordinary constant and $\xi$ is a Grassmann odd
constant. On the other hand, with this $f$, our equations
(\ref{b2}) yield
\[
g_{i,t}-g_{i,xx}+2kg_{i,x}=0,
\]
and
\[
f({\cal D}-\xi-\theta k)(g_{i,t}-g_{i,xx}+2kg_{i,x})=0.
\]
Therefore we can choose
\[
g_i=\mbox{e}^{k_ix+k_i(k_i-2k)t+\theta\xi_i}
\]
as our solutions. Finally, we have
\[
 \tau_1=\mbox{e}^{kx-k^2t+\theta\xi},
 \;\;\tau_2=1+\sum_{i=1}^{N}\mbox{e}^{k_ix+k_i(k_i-2k)t+\theta\xi_i}.
\]

We see that this type of solutions is characterized by the fact
that the coupling constants are zero. Therefore it is the analogy
of the fission and fusion soliton of the BK system
\cite{mm}\cite{skm}(see also \cite{egmm}\cite{cmp}).

\section{B\"{a}cklund Transformation}
B\"{a}cklund transformation is an important ingredient of soliton
theory. It can be useful for solution construction and serves as a
characteristic of integrability for a given system. In this
section, we will derive a bilinear BT for our sTB system. We
follow Nakamura and Hirota \cite{nh} and our results are
summarized in the following

\begin{proposition}
Suppose that $(\tau_1, \tau_2)$ is a solution of the equation
(\ref{BI}), and $(\bar{\tau}_1, \bar{\tau}_2)$ satisfies the
following relations
\begin{eqnarray}\label{bt1}
D_x(\bar{\tau}_1\cdot\tau_2+\tau_1\cdot\bar{\tau}_2)=0,\;\quad\qquad
S(\bar{\tau}_1\cdot\tau_2+\tau_1\cdot\bar{\tau}_2)=0,&&\\
\label{bt2}
D_t(\tau_1\cdot\bar{\tau}_2)-D_x^2(\bar{\tau}_1\cdot\tau_2)=0,\;\;
D_t(\bar{\tau}_1\cdot{\tau}_2)-D_x^2({\tau}_1\cdot\bar{\tau}_2)=0,&&\\
\label{bt3}  SD_t\tau_1\cdot\bar{\tau}_2+{1\over
  2}SD_x^2\tau_1\cdot\bar{\tau}_2-{1\over
  2}SD_x^2\bar{\tau}_1\cdot\tau_2=0,&&\\
\label{bt4} SD_t\bar{\tau}_1\cdot{\tau}_2+{1\over
  2}SD_x^2\bar{\tau}_1\cdot{\tau}_2-{1\over
  2}SD_x^2{\tau}_1\cdot\bar{\tau}_2=0,&&
\end{eqnarray}
 is the another solution of (\ref{BI}).
\end{proposition}
 {\em Proof}: We consider the following
\begin{eqnarray*}
 {\mathbb{ P}}_1&=&D_x\Big[[(D_t+D_{x}^2)\tau_1\cdot\tau_2]
 \cdot\bar{\tau}_1\bar{\tau}_2-\tau_1\tau_2\cdot[(D_t+D_{x}^2)\bar{\tau}_1\cdot\bar{\tau}_2]
 \Big],\\
 \mathbb{P}_2&=&\Big[[S(D_t+D_{x}^2)\tau_1\cdot\tau_2]
 \bar{\tau_1}\bar{\tau}_2+\tau_1\tau_2[S(D_t+D_{x}^2)\bar{\tau}_1\cdot\bar{\tau}_2]
 \Big].
\end{eqnarray*}
We will show that above equations (10-13) imply $\mathbb{P}_1=0$
and $\mathbb{P}_2=0$. The case of $\mathbb{P}_1$ can be verified
as in \cite{nh}, so we will concentrate on $\mathbb{P}_2$. We will
use various bilinear identities which are presented in the
Appendix.
\begin{eqnarray*}
\mathbb{P}_2&=&(SD_t\tau_1\cdot\tau_2)\bar{\tau}_1\bar{\tau}_2+\tau_1\tau_2(SD_t\bar{\tau}_1\cdot\bar{\tau}_2)+
(SD_x^2\tau_1\cdot\tau_2)\bar{\tau}_1\bar{\tau}_2+\tau_1\tau_2(SD_x^2\bar{\tau}_1\cdot\bar{\tau}_2)\\
&\overset{(\ref{A5}),(\ref{A7})}{=}&(SD_t\tau_1\cdot\bar{\tau}_2)\bar{\tau}_1{\tau}_2+\tau_1\bar{\tau}_2(SD_t\bar{\tau}_1\cdot{\tau}_2)
+(S\tau_1\cdot\bar{\tau}_1)(D_t\bar{\tau}_2\cdot\tau_2)+(D_t\tau_1\cdot\bar{\tau}_1)(S\bar{\tau}_2\cdot\tau_2)\\[4pt]
&&+D_x^2\{(S\tau_1\cdot\bar{\tau}_2)\cdot\bar{\tau}_1\tau_2+\tau_1\bar{\tau}_2\cdot(S\bar{\tau}_1\cdot\tau_2)\}
-(S\tau_1\cdot\tau_2)(D_x^2\bar{\tau}_1\cdot\bar{\tau}_2)\\[4pt]
&&-(S\bar{\tau}_1\cdot\bar{\tau}_2)(D_x^2{\tau}_1\cdot{\tau}_2)+
2(SD_x\tau_1\cdot\tau_2)(D_x\bar{\tau_1}\cdot\bar{\tau}_2)+2(SD_x\bar{\tau}_1\cdot\bar{\tau}_2)(D_x{\tau}_1\cdot{\tau}_2)\\[4pt]
&\overset{(\ref{A6})}{=}&(SD_t\tau_1\cdot\bar{\tau}_2)\bar{\tau}_1{\tau}_2+\tau_1\bar{\tau}_2(SD_t\bar{\tau}_1\cdot{\tau}_2)+
(S\tau_1\cdot\tau_2)(D_t\bar{\tau}_2\cdot\bar{\tau}_1)+(D_t\tau_1\cdot\tau_2)(S\bar{\tau}_2\cdot\bar{\tau}_1)\\[4pt]
&&+(S\tau_1\cdot\bar{\tau}_2)(D_t\bar{\tau}_1\cdot{\tau}_2)+(D_t\tau_1\cdot\bar{\tau}_2)(S\bar{\tau}_1\cdot{\tau}_2)
+D_x^2\{(S\tau_1\cdot\bar{\tau}_2)\cdot\bar{\tau}_1\tau_2\\[4pt]
&&+\tau_1\bar{\tau}_2\cdot(S\bar{\tau}_1\cdot\tau_2)\}-
(S\tau_1\cdot\tau_2)(D_x^2\bar{\tau}_2\cdot\bar{\tau}_1)-(D_x^2\tau_1\cdot\tau_2)(S\bar{\tau}_1\cdot\bar{\tau}_2)
\\[4pt]
&&+2(SD_x\tau_1\cdot\tau_2)(D_x\bar{\tau}_1\cdot\bar{\tau}_2)+2(SD_x\bar{\tau}_1\cdot\bar{\tau}_2)(D_x{\tau_1}\cdot{\tau}_2)\\[4pt]
&=&(SD_t\tau_1\cdot\bar{\tau}_2)\bar{\tau}_1{\tau}_2+\tau_1\bar{\tau}_2(SD_t\bar{\tau}_1\cdot{\tau}_2)+
(S\tau_1\cdot\bar{\tau}_2)(D_t\bar{\tau}_1\cdot{\tau}_2)+(D_t\tau_1\cdot\bar{\tau}_2)(S\bar{\tau}_1\cdot{\tau}_2)\\[4pt]
&&+D_x^2\{(S\tau_1\cdot\bar{\tau}_2)\cdot\bar{\tau}_1\tau_2+\tau_1\bar{\tau}_2\cdot(S\bar{\tau}_1\cdot\tau_2)\}+
2(SD_x\tau_1\cdot\tau_2)(D_x\bar{\tau_1}\cdot\bar{\tau}_2)\\[4pt]
&&+2(SD_x\bar{\tau_1}\cdot\bar{\tau}_2)(D_x{\tau}_1\cdot{\tau}_2)\\
&\overset{(\ref{A8})}{=}&\left[(SD_t+{1\over
2}SD_x^2)\tau_1\cdot\bar{\tau}_2\right]\bar{\tau_1}{\tau}_2+\tau_1\bar{\tau}_2\left[(SD_t+{1\over
2}SD_x^2)\bar{\tau}_1\cdot{\tau}_2\right]
\\
&&+(S\tau_1\cdot\bar{\tau}_2)\left[(D_t+{1\over
2}D_x^2)\bar{\tau}_1\cdot{\tau}_2\right]+\left[(D_t+{1\over
2}D_x^2)\tau_1\cdot\bar{\tau}_2\right](S\bar{\tau}_1\cdot{\tau}_2)\\
&&+{1\over
2}D_x^2\{(S\tau_1\cdot\bar{\tau}_2)\cdot\bar{\tau}_1\tau_2+\tau_1\bar{\tau}_2\cdot(S\bar{\tau}_1\cdot\tau_2)\}-
(SD_x\bar{\tau}_2\cdot\tau_1)(
D_x\tau_2\cdot\bar{\tau}_1)\\
&&-(SD_x{\tau}_2\cdot\bar{\tau}_1)
(D_x\bar{\tau}_2\cdot{\tau}_1)\\
&\overset{(\ref{bt2}-\ref{bt4})}{=}&{1\over
2}(SD_x^2\bar{\tau}_1\cdot\tau_2)\bar{\tau}_1\tau_2+{1\over
2}\tau_1\bar{\tau}_2(SD_x^2{\tau}_1\cdot\bar{\tau}_2)+(S\tau_1\cdot\bar{\tau}_2)\left[D_x^2\tau_1\cdot\bar{\tau}_2+{1\over
2}D_x^2\bar{\tau}_1\cdot\tau_2\right]\\
&&+(S\bar{\tau}_1\cdot{\tau}_2)\left[D_x^2\bar{\tau}_1\cdot{\tau}_2+{1\over
2}D_x^2{\tau_1}\cdot\bar{\tau}_2\right]+{1\over
2}D_x^2\{(S\tau_1\cdot\bar{\tau}_2)\cdot\bar{\tau}_1\tau_2+\tau_1\bar{\tau}_2\cdot(S\bar{\tau}_1\cdot\tau_2)\}\\
&&-(SD_x\bar{\tau}_2\cdot\tau_1)(
D_x\tau_2\cdot\bar{\tau}_1)-(SD_x{\tau}_2\cdot\bar{\tau}_1)
(D_x\bar{\tau}_2\cdot{\tau}_1)\\
 &=&{1\over
2}(SD_x^2\bar{\tau}_1\cdot\tau_2)\bar{\tau}_1\tau_2+{1\over
2}\tau_1\bar{\tau}_2(SD_x^2{\tau_1}\cdot\bar{\tau}_2)+\left[
(S\tau_1\cdot\bar{\tau}_2)+{1\over
2}(S\bar{\tau}_1\cdot{\tau}_2)\right](D_x^2\tau_1\cdot\bar{\tau}_2)\\
&&+\left[ (S\bar{\tau_1}\cdot{\tau}_2)+{1\over
2}(S{\tau_1}\cdot\bar{\tau}_2)\right](D_x^2\bar{\tau}_1\cdot{\tau}_2)+{1\over
2}D_x^2\{(S\tau_1\cdot\bar{\tau}_2)\cdot\bar{\tau}_1\tau_2+\tau_1\bar{\tau}_2\cdot(S\bar{\tau}_1\cdot\tau_2)\}\\
&&+(SD_x{\tau}_1\cdot\bar{\tau}_2)(
D_x\bar{\tau}_1\cdot{\tau}_2)+(SD_x\bar{\tau}_1\cdot{\tau}_2)
(D_x{\tau}_1\cdot\bar{\tau}_2)
\end{eqnarray*}
\begin{eqnarray*}
&\overset{(\ref{bt1})}{=}&{1\over
2}\Big[(SD_x^2\bar{\tau}_1\cdot\tau_2)\bar{\tau}_1\tau_2+(S\bar{\tau}_1\cdot\tau_2)(D_x^2\bar{\tau}_1\cdot\tau_2)-
D_x^2[(S\bar{\tau}_1\cdot\tau_2)\cdot\bar{\tau}_1\tau_2]\\
&&-2(SD_x\bar{\tau}_1\cdot{\tau}_2)( D_x\bar{\tau}_1\cdot{\tau}_2)
\Big]+{1\over
2}\Big[(SD_x^2{\tau}_1\cdot\bar{\tau}_2){\tau}_1\bar{\tau}_2+(S{\tau}_1\cdot\bar{\tau}_2)(D_x^2{\tau}_1\cdot\bar{\tau}_2)
\\
&&-D_x^2[(S{\tau}_1\cdot\bar{\tau}_2)\cdot{\tau}_1\bar{\tau}_2]-2(SD_x{\tau}_1\cdot\bar{\tau}_2)(
D_x{\tau}_1\cdot\bar{\tau}_2) \Big]\\
&\overset{(\ref{A9})}{=}&0,
\end{eqnarray*}
thus our proposition is proved.

\section{Discussion}
We constructed the bilinear form for the sTB system, then we
presented a class of solutions. This kind of solutions is the
generalization of the fusion and fission solutions of the cB
equation \cite{mm}\cite{skm} (see \cite{lou} for other systems
which possess this sort of solutions). Apart from the fusion and
fission solitons, there is another type of solitons, which was
constructed by Kaup \cite{kaup} in the framework of inverse
scattering transformation and by Hirota and Satsuma using bilinear
formulism \cite{hs}. However, the present version of
bilinearization does not apparently allow one to calculate this
sort of solution in the supersymmetric case. It would be
interesting to find out if such soliton solutions exist or not in
the supersymmetric case.

\vspace{10pt}
 {\bf Acknowledgements} We should like to thank Xing-Biao Hu for interesting
 discussions. It is pleasure to thank the anonymous referees for
 pointing out many typos of early version of the paper.
    This work is supported in part by NNSFC under the
grant number 10231050 and the Ministry of Education of China.

\vspace{15pt}
\renewcommand{\theequation}{A.\arabic{equation}}
\setcounter{equation}{0}
\section*{Appendix: Some Bilinear Identities}
 In this Appendix, we list the relevant bilinear identities, which can be proved
 directly. Here $a$, $b$, $c$ and $d$ are arbitrary even functions
 of the independent variable $x$, $t$ and $\theta$.
\begin{eqnarray}\label{A5}
(SD_ta\cdot b)cd+ab(SD_tc\cdot d)&=&(SD_ta\cdot
d)cb+ad(SD_tc\cdot b)\nonumber\\
&&+(Sa\cdot c)(D_td\cdot b)+(Sd\cdot b)(D_ta\cdot c),
\end{eqnarray}
\begin{eqnarray}\label{A6}
(Sa\cdot b)(D_t c\cdot d)+(D_ta\cdot b)(S c\cdot d)&=&(Sa\cdot
d)(D_t c\cdot b)+(D_ta\cdot d)(S c\cdot b)\nonumber\\
&&+(Sa\cdot c)(D_t b\cdot d)+(D_ta\cdot c)(S b\cdot d),
\end{eqnarray}
\begin{eqnarray}\label{A7}
(SD_x^2a\cdot b)cd+ab(SD_x^2c\cdot d)&=&D_x^2\{(Sa\cdot
d)\cdot cb+ad\cdot (Sc\cdot b)\}{}\nonumber\\
&&{}-(Sa\cdot b)D_x^2(c\cdot d)-(Sc\cdot d)D_x^2(a\cdot
b)\nonumber\\
&&+ 2(SD_xa\cdot b)D(c\cdot d)+2(SD_x c\cdot d)(D_xa\cdot b){},
 \end{eqnarray}
\begin{eqnarray}\label{A8}
(SD_x^2a\cdot b)cd+ab(SD_x^2c\cdot d)&=& D_x^2\{(Sa\cdot b)\cdot
cd+ab\cdot (Sc\cdot d)\}{}\nonumber\\
&&-(Sa\cdot b)(D_x^2c\cdot d)-(Sc\cdot d)(D_x^2a\cdot b)\nonumber\\
&&-4(SD_xb\cdot c)(D_xd\cdot a)-4(SD_xd\cdot a)(D_xb\cdot c)\nonumber\\
&&-2(SD_xa\cdot b)(D_xc\cdot d)-2(SD_xc\cdot d)(D_xa\cdot b),
\end{eqnarray}
\begin{eqnarray}\label{A9}
D_x^2[(Sa\cdot b)\cdot ab]-(SD_x^2a\cdot b)ab-(Sa\cdot
b)D_x^2(a\cdot b)+2(SD_xa\cdot b)(D_xa\cdot b)=0.
\end{eqnarray}

\end{document}